\documentclass[12pt]{iopart}

\usepackage{epsfig}
\usepackage{iopams}
\usepackage{graphicx}

\begin{document}

\title{Random neighbour model for yielding}

\author{F. Dalton$^{1}$, A. Petri$^{1,2}$ and G. Pontuale$^{1}$}
\address{$^{1}$Consiglio Nazionale delle Ricerche, Istituto dei Sistemi Complessi, via del Fosso del Cavaliere 100, 00133 Roma - Italy}
\address{$^{2}$Dipartmento di Fisica, Sapienza Universit\`a, P.le A. Moro 5, 00189 Roma - Italy}
\ead{\mailto{fergal.dalton@isc.cnr.it}, \mailto{alberto.petri@isc.cnr.it}, \mailto{giorgio.pontuale@isc.cnr.it}}
\date{\today}

\begin{abstract}

We introduce a model for yielding, inspired by fracture models and the failure of a sheared granular medium in which the
applied shear is resisted by self-organized force chains.
The force chains in the granular medium (GM) are considered as a bundle of fibres of finite strength amongst
which stress is randomly redistributed after any other fibre breaks under excessive load.
The model provides an exponential distribution of the internal stress and a log-normal shaped distribution of failure
 stress, in agreement with experimental observations.
The model displays critical behaviour which approaches mean field as the number of random neighbours $k$ becomes large and also displays
a failure strength which remains finite in the limit of infinite size.
From comparison with different models it is argued that this is an effect of uncorrelation. 
All these macroscopic properties appear statistically stable with respect to the choice of the chains' initial strength distribution.
The investigated model is relevant for all systems in which some generic external load or
pressure is borne by a number of units, independent of one another except when
failure of a unit causes load transfer to some random choice of neighbouring units.

\end{abstract}
\submitto{Journal of Statistics and Mechanics: theory and experiment}
\maketitle

\section{Introduction}
\label{sec:intro}
Many different processes, from stress propagation in a heterogeneous medium to the services for client satisfaction involve
redistribution of loads when some units break down, potentially
triggering an avalanche of failures and causing catastrophic breakdown of the entire system.
Other examples might be  data centers which supply information over some
communications infrastructure and electrical power supply from power stations.
Such generic non-linear and irreversible failure processes, where the internal dynamics can be highly susceptible to
external perturbations, constitute a general feature of yielding processes and  may be observed in many other different
contexts and systems such as fracture, plastic deformation, granular slip, economic crashes, etc.
This class of phenomena generally exhibits critical behaviour with relevant
characterizing features such as finite yield points and broad avalanche amplitude
distributions, as well as a highly inhomogeneous internal load distribution over the
load-bearing units.

In this article, we propose a simple model for yielding that is general in purpose but 
essentially inspired by the case
of a Granular Medium (GM) under shear stress.
The distribution of stress within a GM is a characteristic feature
at variance with fluids; it is neither isotropic nor homogeneous due to
the presence of force chains
\cite{drescher72,behringer91}, highly singular random paths
which bear most of the load and screen the majority of grains from stress.
In real GMs, the failure of a force chain triggers stress transfer processes
which involve complex spatial structures \cite{vallejo05}
which can also dynamically change under the chain rupture and alter the
redistribution process itself. These cannot easily
be incorporated in any simple model without resorting to molecular dynamics simulations.
The model presented here uses a simplifying approach in which force chains,
each represented by a single number, are the smallest components.
Each chain in the medium is able to support a maximum load, and
as the increasing external stress causes the weakest remaining fibre to break,
its load is redistributed to a number of other surviving chains.  The model shares some
features with the Fibre Bundle Model (FBM) for fractures~\cite{peirce26}, but introduces
a novel random load redistribution mechanism.   For this reason, it shall
henceforth be denoted the Random Fiber Bundle Model (RFBM).
With respect to the sheared granular bed, fibres play the
role of the force chains within the granular material, an analogy already exploited
in \cite{hidalgo02} for compressional stress in which broken chains undergo a restructuring 
which results in a stiffening of the bundle. In the RFBM model presented here, representing brittle shear failure,
breaking force chains transfer their stress to some random selection of force chains.
The model aims to reproduce some macroscopic statistical features of the yield process 
already known from experimental measurements, as follows.

\textit{i)} Experimental studies \cite{liu95,behringer99} have pointed out
 that stress distributions within a granular
medium are predominantly characterized by long exponential tails. This property
seems to be general since it appears in different regimes of
granular mechanics ranging from static assembly \cite{liu95} to
dynamic shear \cite{behringer05}.  A simple explanation for the static case
is given by the so called \emph{q-model} \cite{coppersmith96} which, although imprecise in some aspects
\cite{marconi00}, produces robust exponential tails.  In the q-model,
grains arranged in horizontal layers discharge random proportions of
their load on grains in the layer below.  The model and its developments have been considered
and discussed in several works \cite{socolar98,cates98,marconi00,rajesh00,hern01,lewandowska01},
from which the idea emerges that random load transfer
can produce long exponential tails as a stable characteristic.
The RFBM model proposed here allows force chains to irreversibly break, but retains this random rule for load redistribution,
by randomly selecting which of the remaining fibres should bear the load of the breaking fibre.
The randomness of the choice may be interpreted as an expression of our inherent ignorance
about the real stress and complex spatial force chain correlations.

\textit{ii)} Another characteristic of the shear yield process is that the
force required to trigger a slip in a
sheared granular material, as well the force required to sustain
steady motion, are statistical quantities that have been shown to accurately follow log-normal
shaped distributions \cite{dalton05,petri08}. This
feature does not seem to be exclusive to granular
yield and has in fact been observed in solid-on-solid
\cite{johansen93} and fibre-on-fibre \cite{briscoe85} friction not to mention
more generic yielding phenomena ranging from friction and wear \cite{steele08}
to cohesion and bearing capacity of soils \cite{fenton00} and the strength of
wood structures \cite{kazum99}. Moreover, log-normal shaped distributions
may also be produced by a variety of model systems ranging
from fracture \cite{alava06} to dislocation \cite{beato},
as well as magnetic and self organized systems \cite{bramwell02}.

\textit{iii)} Finally,  many yield processes display critical features. Typical examples are the depinning transition of elastic manifolds \cite{chakrabarti97} and fracture processes \cite{chakrabarti97}.  Granular dynamics proper has been shown to display critical dynamics in the chaotic stick-slip motion phase \cite{baldassarri06}.


It will be shown that the RFBM reproduces all the main features described above, i.e. \textit{i)}
the exponential distribution of internal loads, \textit{ii)} a non Gaussian (log-normal shaped) probability  distribution of failure loads and \textit{iii)} a critical dynamics characterized by a mean field exponent and a cut-off, the divergence with size of which depends on the number of abssorbing chains.  In addition, it will be seen that the model exhibits a finite yield strength even in the infinite system limit. Although this latter case is difficult (sic!) to test experimentally, it is at least clear that it holds in finite size cases \cite{dalton05,lu07,mills08}. From the comparison with models having different geometries and load sharing rules we will conclude that such a feature is related to the absence of spatial correlation due to the random neighbour allocation.


This article is organized as follows:
Section~\ref{sec:details} describes the RFBM and recalls some related models in order to compare their properties.
The distribution of the internal stress is investigated  in section~\ref{sec:internalloads} and section~\ref{sec:yieldpoint}
discusses the statistical properties of the yield point stress.  Section~\ref{sec:critical} shows the critical aspect of the process,
drawing a careful comparison with those of the usual FBM.  Concluding remarks are drawn in section~\ref{sec:summary}.

\section{Model details and features}
\label{sec:details}
The dynamics of the model discussed in the present work is to some extent based on the Fibre-Bundle Model (FBM), a
simple but revealing approach to the study of failure originally
introduced by Peirce \cite{peirce26} to explore the properties of
textiles and subsequently extensively explored
\cite{hidalgo02,daniels45,suh70,sornette89,hansen92,kloster97,pradhan03}.
The FBM consists of a bundle of $N$ harmonic fibres loaded in parallel
by a slowly increasing external load $F$.  Each fibre has a different breaking
strength $s_i$ drawn from some identical initial distribution $t(s)$.
As the  external load is increased, fibres whose load exceeds their strength
break, and redistribute their load equally on each of the remaining intact fibres $R$.
("global load sharing" law). Every intact fibre therefore,
carries the same load and in this respect it can be considered the
simplest mean field model for fracture.

As in the FBM we define the RFBM by a set of harmonic
fibres each having identical elastic modulus but being able to support a different random
maximum failure load $s_i$ extracted from some probability distribution $t(s)$.
The bundle is subjected to an external load
$F$, initially zero, which is then increased so as to break only the weakest fibre
which, in the RFBM, then redistributes its load on a fixed number $k$ of different randomly chosen intact fibres,
any of which may now exceed their capacity and break in turn.
Thus we may observe an ``avalanche'' of breaking
fibres. After the avalanche (if any) the external load is increased again so
as to break only the new weakest fibre (this corresponds to the limit of adiabatic forcing).
We investigate several variants of this basic RFBM.
The case $k=1$, in which a breaking fibre passes its entire load to a single other fibre;
the case $k=2$ where the load is redistributed in \textit{i)}
equal proportions (ES) or \textit{ii)} randomly chosen proportions $p$ and $q=1-p$ (PQS) among the two randomly chosen
fibres ($p$ is newly extracted at the breaking of each fibre). For the cases  $k=3,4$ some main properties are discussed.
The rules of the RFBM bear some similarities with the conservative limit of the random
neighbour sand pile model \cite{lise96,broker97} except that here the threshold of each fibre (site)
is drawn from a random distribution and fibre breaking is irreversible.
However, like the sandpile model, every intact fibre bears a different load which depends on the fibre's history.

We define the stress as $\sigma=\sum \tau_i / N = F/N$, where $\tau_i$ is the load on the $i^\textrm{th}$ fibre
and $N$ is the initial number of fibres. The yield stress $\sigma_c$ then, is $\sigma$ evaluated at
the beginning of the final avalanche which ruptures the bundle as a whole.
We define the reduced bundle load as $f=\sigma/\sigma_c$ which
varies from zero to one as the simulation proceeds,
and permits the comparison of results from different realizations.

In the original FBM, $\sigma_c$ is a random number for small
$N$,   with a skewed  distribution which
tends to a Gaussian with increasing $N$ and a $\delta$ function for $N=\infty$.
A non-mean field version of the FBM also exists, with fibres placed on a regular lattice, and
discharging to their nearest neighbours only (Local Sharing rule - LFBM) \cite{zhang96,kloster97}. The behaviour
of this model has also been recently investigated on different types
of complex networks (CFBM) \cite{kim05}.
For comparison we will also present and discuss some features and  results from these variations.
It will be seen that the RFBM  exhibits behaviour which in some sense places  it between the FBM and LFBM.
Except where otherwise specified, all simulations are performed on systems with $N=10,000$ fibres.

\section{The internal distribution of loads}
\label{sec:internalloads}

As noted in the introduction, our main motivation for devising and investigating the RFBM
 is to attempt a description of the yield of a sheared granular bed.
 In doing so, one reduces the granular force chains to simple fibres, characterized by a single
variable $s$, their ability to bear stress.  It is known that a GM under either static or
dynamic loading displays an exponential distribution of internal
forces \cite{liu95,coppersmith96,socolar98,behringer99,cates98,marconi00,rajesh00,hern01,lewandowska01,behringer05},
a property generally encountered in disordered systems \cite{makse00,brujic03,hern01}.
We have computed therefore the internal load distribution $p(\tau)$ as a function
of the reduced bundle load $f=\sigma/\sigma_c$.

\begin{figure}
\begin{center}
\includegraphics[height=11cm,angle=270]{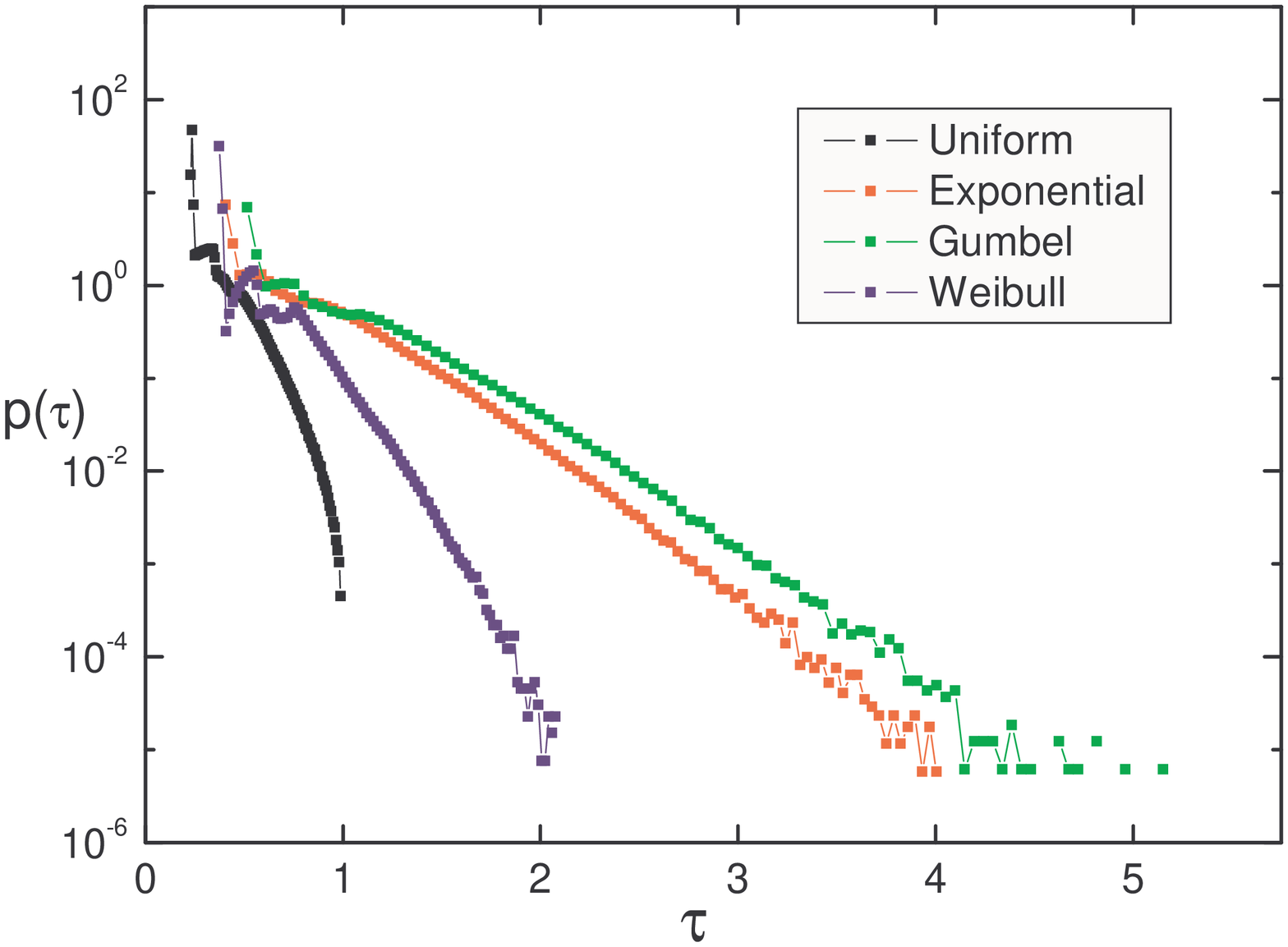}
\end{center}
\caption{\label{fig1} The distribution of the load per fibre $p(\tau)$ for the RFBM ES
($k=2$) at the yield point $f=1$. $p(\tau)$  shows
 stable long exponential tails for several different
probability distributions of the fibre strength $t(s)$ (see Appendix for the distributions details).}
\end{figure}

\begin{figure}
\begin{center}
\includegraphics[height=11cm,angle=270]{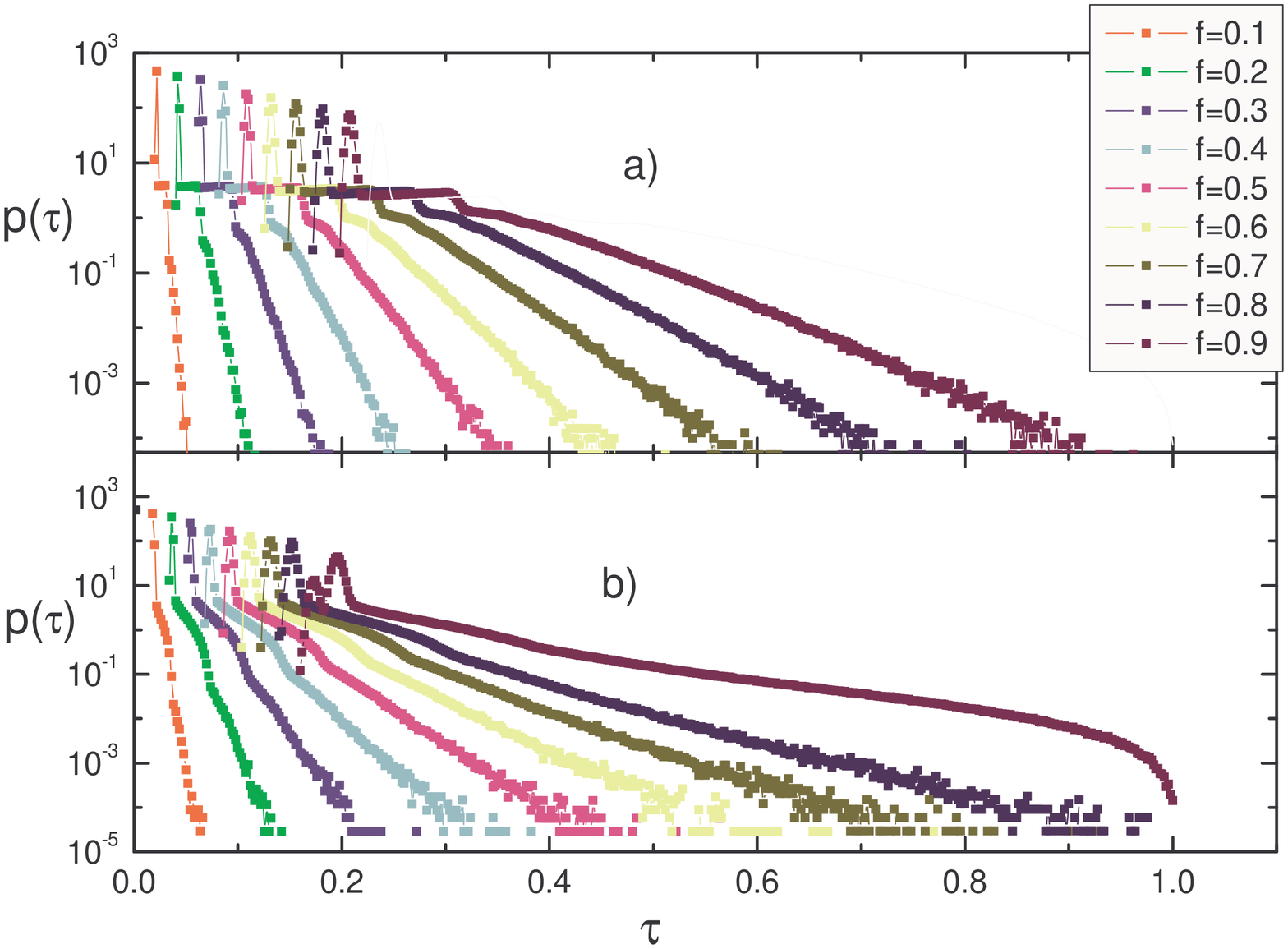}
\end{center}
\caption{\label{fig2} Distribution of the load per fibre $p(\tau)$ in the, $k=2$, RFBM for
an exponential distribution of fibre strenght: $t(s)=exp(-s), \, \, s \ge 0$ at increasing values of $f$: a) Equal sharing (ES);  b) Random sharing (PQS). Exponential tails in $p(\tau)$ are established early in the system's evolution.}
\end{figure}

In figure~\ref{fig1} we demonstrate this distribution at the yield point $f=1$ for four different choices of the fibres failure threshold distribution $t(s)$
(uniform, exponential, Gumbel and Weibull, details in the appendix),
with the random neighbour number $k=2$ and equal sharing (ES).
The figure shows that the RFBM exhibits robust exponential tails in the load distribution.
Since initially $f=0$ and the load on each fibre is zero, this distribution builds up according as the
simulation proceeds and the external load is increased.
Figure~\ref{fig2} illustrates the evolution of the internal load distribution $p(\tau)$ for increasing values of $f$ and 
$t(s)$ exponential. The ES case is shown in a), whereas b) shows results for random sharing on $k=2$ fibres (PQS).
It is clear that the exponential tail in the distribution of internal forces is achieved quite
early in the evolution of the system, certainly long before yielding occurs.
All variants of the model investigated display exponential tails similar to the $k=2$, in particular
for $k=1,3,4$, and using uniform  $t(s)$ instead of exponential.

\begin{figure}
\begin{center}
\includegraphics[height=13cm,angle=270]{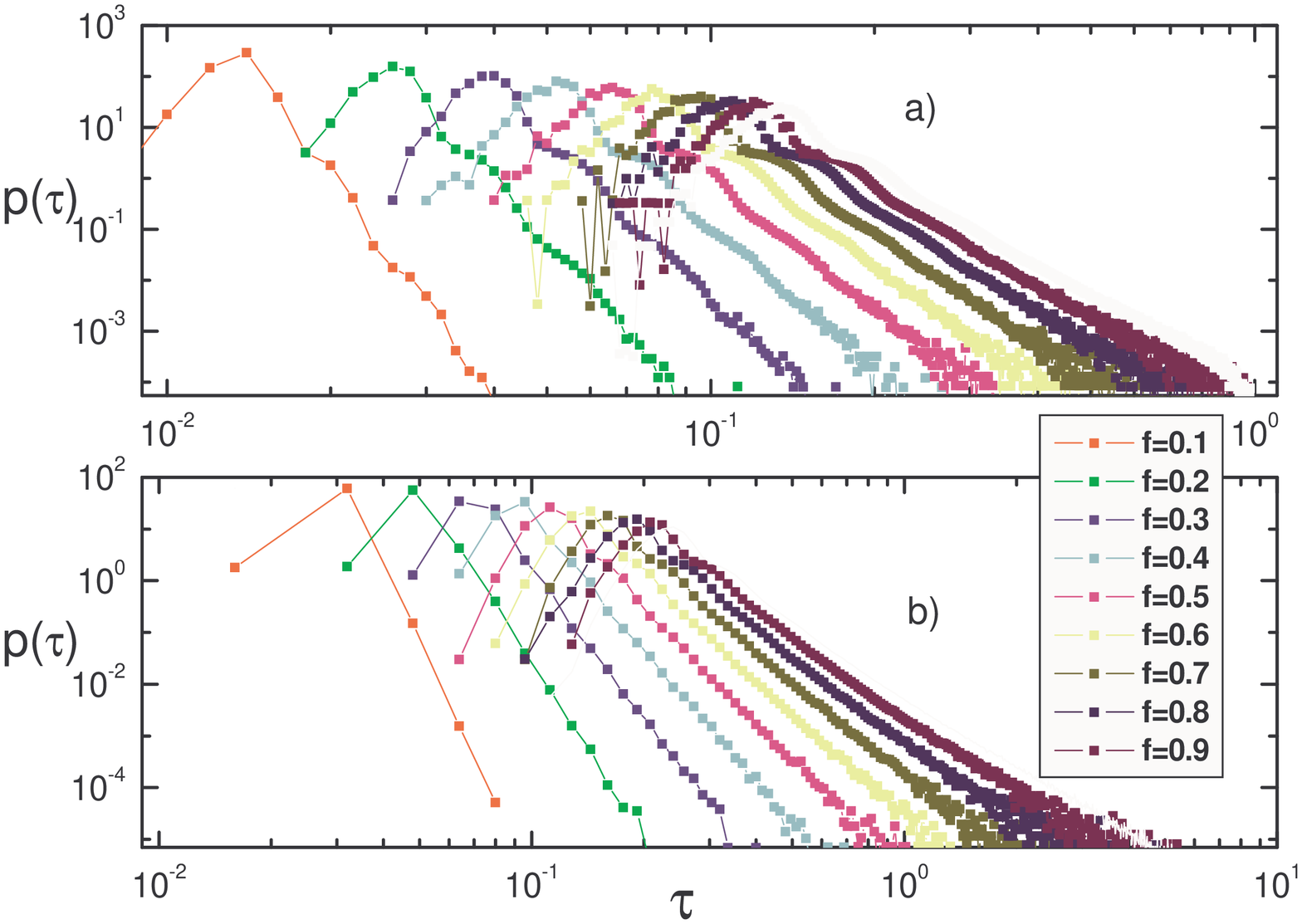}
\end{center}
\caption{\label{fig3}Distribution of the load per fibre $p(\tau)$ at different reduced external loads $f$ in a 1D
model with local load sharing (LFBM) for fibre strength distributions $t(s)$: a) uniform; b) exponential. In
this case  $p(\tau)$ is algebraic in character.}
\end{figure}

From the above cases we conclude that exponential tails in the
internal distribution of loads are robust features of the model, and the generality of
results when changing $t(s)$ and $k$ suggests that the random load redistribution
property of the model is a general mechanism naturally leading to the
exponential load distribution \cite{rajesh00}, analogous to that of the $q$-model.
As a test we have also computed  $p(\tau)$ for the LFBM in a 1D periodic geometry, in which
the load of a broken fibre is transferred to $k=2$ neighbouring fibres.
The resulting $p(\tau)$ at different values of the external load are shown in
figure \ref{fig3}. In both cases investigated ($t(s)$ uniform and exponential), the internal load distribution
displays long algebraic tails.   This, in itself,
is a remarkable feature of the LFBM and merits further study.
From these results and those related to the q-model and similar models \cite{rajesh00}, we conlcude that the
exponential distribution of loads obtained above is specific to random correlations. This, of course, does not exclude the possibility that other mechanisms
may also generate exponential distributions of loads, though we suggest that any other such mechanism will hardly be simpler that that presented here.

\section{The yield point}
\label{sec:yieldpoint}

In this section we investigate the behaviour of the RFBM at the system's global yield point, varying
a number of parameters such as the bundle size $N$ and the fibre strength distribution $t(s)$.  We
compute in particular  $g(\sigma_c)$, the probability that the system yields at external stress $\sigma_c$. Among other
results we find that: \textit{i)} the dependence of the shape of $g(\sigma_c)$ on the fibre strength distribution $t(s)$ is very weak;
\textit{ii)} for not too large $N$, $g(\sigma_c)$ has a log-normal shape and \textit{iii)} the mean, variance and skewness of $g(\sigma_c)$ decreases algebraically with the system size.

 We consider initially the case $k=2$.Experimental measurements of the yield point of a sheared granular medium have been performed extensively in \cite{dalton05,petri08}, and have demonstrated the distribution to be skewed and far from  Gaussian. It seems remarkable
that it is also well fitted by a log-normal distribution (or similar)~\cite{dalton05}.
Such a distribution seems to be a generic signature of the strong correlations introduced into
the system by the force chains \cite{petri08}, and some authors have resorted to universality as some underlying feature
for such observations \cite{bramwell02}. The RFBM model outlined here produces a yield stress distribution $g(\sigma_c)$ wich is also very well approximated by a log-normal:
\begin{equation}
\label{eq-lognormal}
g(\sigma_c) = \frac{1}{\sqrt{2\pi}\left|\sigma_c-\sigma_0\right|\delta_{\ln}}
\exp\left[-\left(\frac{\ln(\sigma_c-\sigma_0)-\mu_{\ln}} {\sqrt{2}\delta_{\ln}}\right)^2\right]
\end{equation}
where $\mu_{\ln}$ and $\delta_{\ln}$ are the mean and standard deviation of $\ln(\sigma_c - \sigma_0)$
and $\sigma_0$ is a rigidity threshold which imposes a lower bound to the minimum yield stress
obtainable in any finite system realization.

\begin{figure}
\begin{center}
\includegraphics[height=11cm,angle=270]{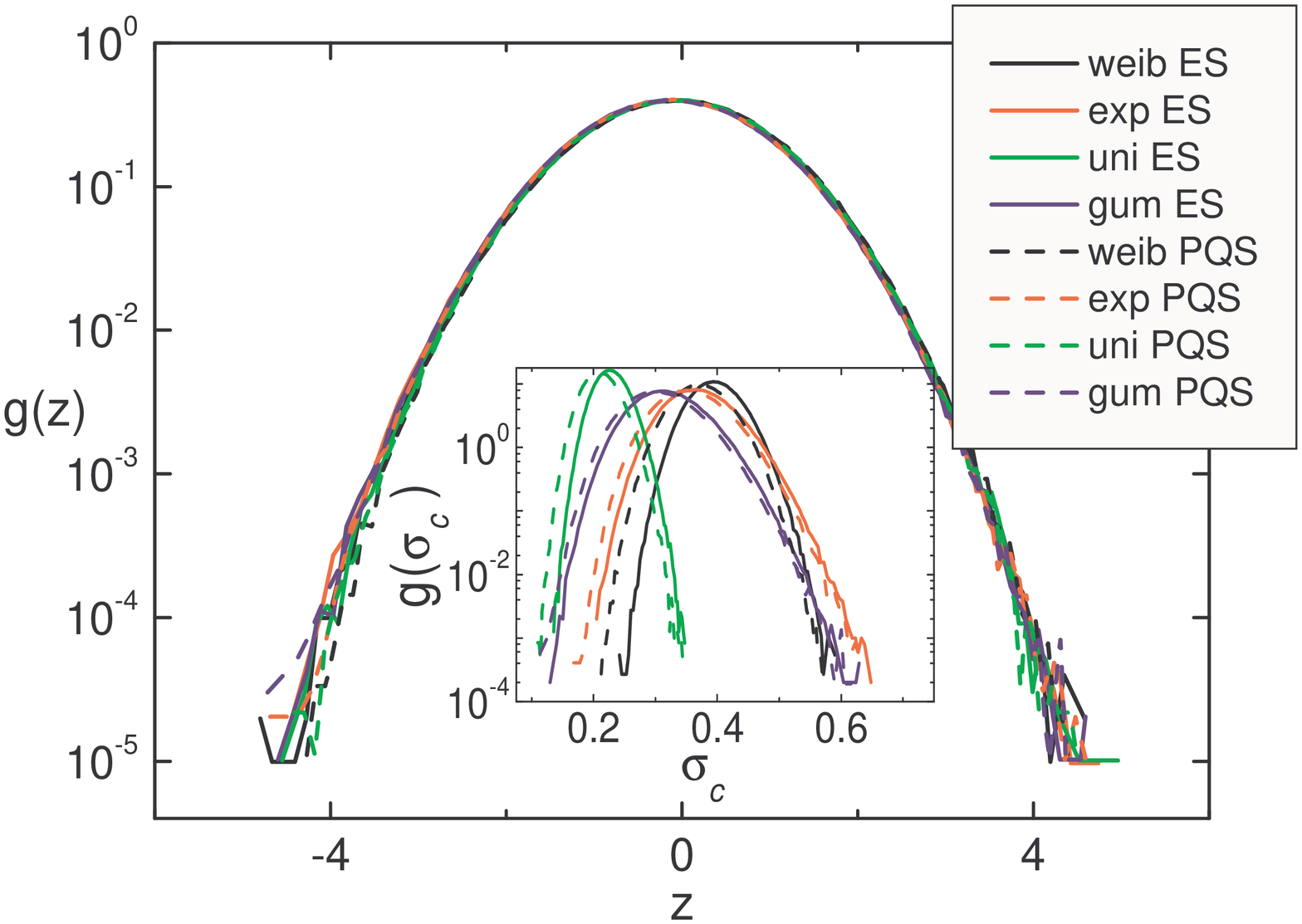}
\end{center}
\caption{\label{fig4} Distribution of the yield strength $z$ in the standardized variable (see text)
for different fibre strength distributions $t(s)$:  uniform,
exponential, Gumbel and Weibull (see Appendix).
The distribution $g(z)$ appears robust with respect to the choice of $t(s)$. Continuous and dashed lines refer to ES and PQS respectively.
The inset displays the same results with non-standardized variables; PQS is seen to be slightly but consistently
weaker than ES.}

\end{figure}

Figure \ref{fig4} shows the  pdf $g(\sigma_c)$, for $N=100$ fibres and four different choices
of the fibres' initial strength distribution $t(s)$: uniform, exponential, Gumbell and Weibull
(see the appendix for details on the precise forms of $t(s)$). Both ES (continuous lines) and
PQS (dashed lines) are shown,  each result averaging over $10^6$ realisations.
When a standardized variable is adopted $z=\left(\ln(\sigma_c-\sigma_0)-\mu_{\ln}\right)/\delta_{\ln}$
all curves collapse to the same normal curve, showing that $g(\sigma_c)$ is not strongly affected by
the choice of fibre strength distribution $t(s)$.
For very low $N$, therefore, $g(\sigma_c)$ is far from Gaussian \cite{daniels45}, becoming Gaussian
only for sufficiently large $N$.
This also follows from the expression for $z$ since: $\mu_{\ln} = \langle\ln(\sigma_c-\sigma_0)\rangle = \ln(\tilde\mu)$ where
$\tilde\mu=\sqrt[n]{\prod(\sigma_c-\sigma_0)}$ is the geometric mean of $(\sigma_c-\sigma_0)$ over $n$ realisations.
As is the case with large $N$, the distributions is narrow, and the geometric mean approximates
the arithmetic mean $\tilde\mu\to\langle\sigma_c-\sigma_0\rangle=\langle\sigma_c\rangle-\sigma_0$ and so
\begin{equation}
z = \frac{1}{\delta_{\ln}}\left[\ln(\sigma_c-\sigma_0)-\ln(\langle\sigma_c\rangle-\sigma_0)\right]
 = \frac{1}{\delta_{\ln}}\ln\left(1+\frac{\sigma_c-\langle\sigma_c\rangle}{\langle\sigma_c\rangle-\sigma_0}\right)
\end{equation}
Thus, if $\frac{\sigma_c-\langle\sigma_c\rangle}{\langle\sigma_c\rangle-\sigma_0} \ll 1$ (i.e.\ the distribution is narrow) then $g(\sigma_c)$ approximates a Gaussian
centered at $\langle\sigma_c\rangle$ with standard deviation $\Delta=\left[\langle\sigma_c\rangle-\sigma_0\right]\delta_{\ln}$:
\begin{equation}
g(\sigma_c) = \frac{1}{\sqrt{2\pi} \Delta}
	\exp\left[-\left(\frac{\sigma_c-\langle\sigma_c\rangle}{\sqrt{2}\Delta}\right)^2\right]
\end{equation}

\begin{figure}
\begin{center}
\includegraphics[height=11cm,angle=270]{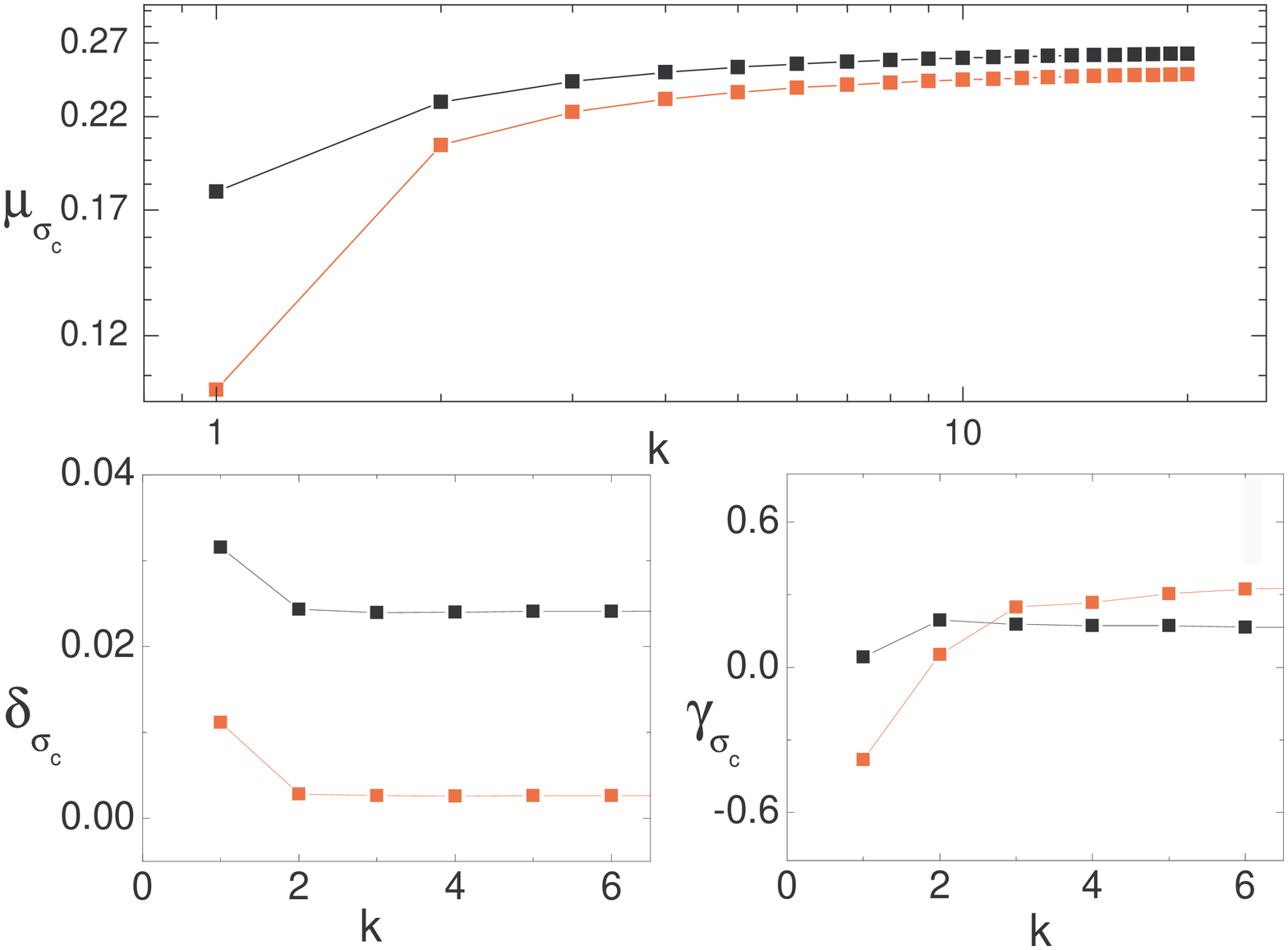}
\end{center}
 \caption{\label{fig5}
Variation of the failure distributions parameters with the random neighbour number $k$ for $N=100$ (black) or $N=10000$ (red) in the ES case. 
a) The mean  b) the standard deviation and c) the skewness of $g(\sigma_c)$.  As the number of neighbours increases,
these parameters rapidly approach the mean field values.
}
\end{figure}

We study now the size dependence of the mean $\mu_{\sigma_c}$, the standard deviation $\delta_{\sigma_c}$ and
the skewness $\gamma_{\sigma_c}$ of the yield point distribution $g(\sigma_c)$, for sizes $N$ from $10^2$ to $10^6$ as
shown in figure~\ref{fig5}.  The ES and PQS cases are shown and compared with the FBM case.
We focus on the case of a uniform $t(s)$ which is more commonly adopted for the
FBM and observe that good power-law fits may be obtained:
\begin{eqnarray}
\label{eq:phichipsi}
\eqalign{
\mu_{\sigma_c} - & \sigma_\infty \propto N^{-\phi} \cr
  & \delta_{\sigma_c} \propto N^{-\chi}\cr
& \gamma_{\sigma_c} \propto N^{-\psi}}
\end{eqnarray} 
where $\sigma_{\infty}$ indicates a non-zero failure load per fibre for an infinite bundle $N=\infty$.

The scaling of $\mu_{\sigma_c}$ in figure~\ref{fig5} a)
shows that an algebraic scaling is common to the FBM (global sharing) and retains
a finite strength $\sigma_\infty$ as $N\rightarrow \infty$.
This is in constrast to the LFBM where, for large $N$, one obtains instead $\mu_{\sigma_{c}} \simeq 1/\ln(N)$~\cite{alava06}, i.e.\ a vanishing 
strength limit for $N\to\infty$.
Despite the local sharing rule in CFBM models, however, a finite strength limit is again retained~\cite{kim05} as in the RFBM; this
may be due to the random network topology that destroys spatial correlations.  Thus the requirement for non-zero $\sigma_\infty$ appears to be an 
absence of spatial correlations, due either to annealed randomness (RFBM) or quenched randomness (CFBM).
Notably, all variants of the RFBM studied possess a
systematically weaker rupture threshold than the FBM, as they tend to concentrate the external load
on some subset of fibres.

Figures \ref{fig5} b) and c)  show the scaling of  the standard deviation and skewness of $g(\sigma_c)$
for sizes $N$ from $10^2$ to $10^6$ fibres.  The skewness vanishes for $N\to\infty$ and the distribution tends to a Gaussian,
consistent with the FBM (also shown) though the RFBM (both ES and PQS), for increasing $N$, gives rise
to symmetric $g(\sigma_c)$  somewhat later than the  FBM model.
The standard deviation $\delta_{\sigma_c} \to 0$ in the limit $N \to \infty$, as expected,  reducing the distribution $g(\sigma_c)$
to a delta-function at $\sigma_\infty$. Scaling exponents appearing in (\refname{eq:phichipsi}) for the models investigated are recorded in table~\ref{tab:exponents}.

\begin{table}
\caption{\label{tab:exponents}Critical exponents for the parameters of the distribution
of failure stresses $g(\sigma_c)$, as defined in \eref{eq:phichipsi}.}
\begin{indented}
\lineup\item[]\begin{tabular}{lcccc}
\br
Model & $\sigma_{\infty}$ & $\phi$ & $\chi$ & $\psi$\\
\mr
FBM & 0.25 & -0.66 & -0.99 & -1.91 \\
RFBM ES & 0.20 & -0.56 & -0.99 & -1.66 \\
RFBM PQS & 0.17 & -0.5 & -0.89 & -1.30 \\
\br
\end{tabular}
\end{indented}
\end{table}

\begin{figure}
\begin{center}
\includegraphics[height=13cm]{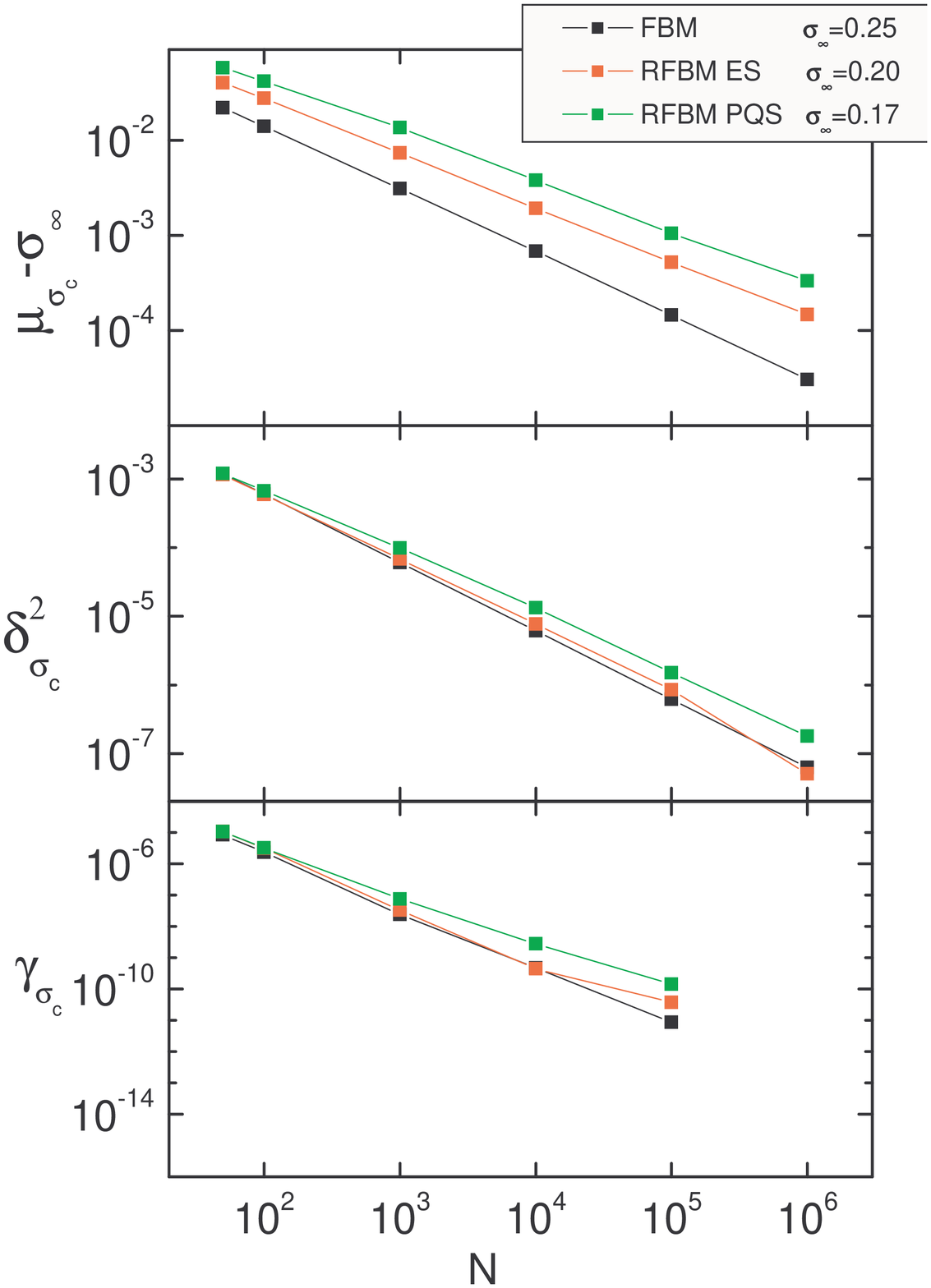}
\end{center}
\caption{\label{fig6} The mean $\mu_{\sigma_c}$, standard deviation $\delta_{\sigma_c}$ and skewness $\gamma_{\sigma_c}$ of the failure stress distribution $g(\sigma_c)$ are shown here as a function of the bundle size $N$ for the FBM and the RFBM ES \& PQS:
a) average, b) variance, c) skewness. There is a clear power-law dependence on $N$ for each of the variables. The skewness for
$N=10^6$  is omitted as it was not possible to obtain a reliable estimation. The exponents are reported in table~\ref{tab:exponents}.}
\end{figure}

The dependence of the mean, standard deviation and skewness of the yield stress distribution $g(\sigma_c)$ on
the number of random neighbours $k$ is shown in figure~\ref{fig6}.  Values are reported for two different choices of the system size, $N=100$ and $N=10000$.  They are shown to rapidly approach their mean-field values as $k$ increases.

\section{Critical features}
\label{sec:critical}

\begin{figure}
\begin{center}
\includegraphics[height=13cm]{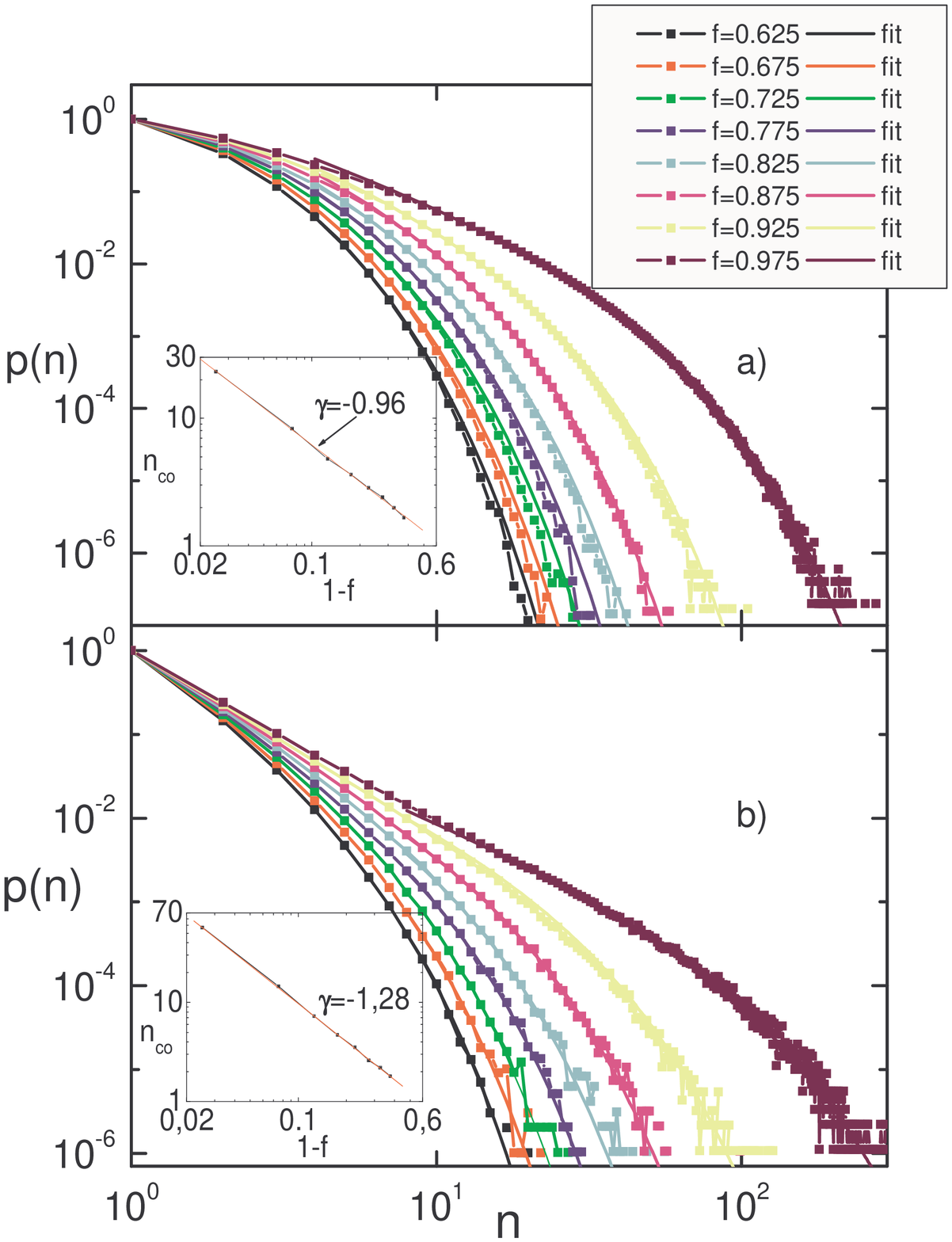}
\end{center}
\caption{\label{fig7}Avalanche size distribution $p(n)$ at several values of $f$ for: a) The FBM (global load sharing, $t(s)$ uniform)
 at several values of the external load $f$, together with the curve fits by (\ref{eq:avalanches}). b) he RFBM ES ($k=2$, uniform $t(s)$).
Insets displays the dependence of the cut-off $n_{co}$ on $f$ according to
(\ref{eq:co}).} 
\end{figure}

One remarkable feature of the FBM is that, during loading, fibres break
in bursts giving rise to a highly intermittent and erratic activity
with no characteristic size.
This kind of activity is actually observed in real fracture experiments (e.g. through the emission of elastic waves),
from microfractures and crystal defects to earthquakes, and is one of
the reasons that this model is the subject of many investigations  \cite{alava06}.
Considering all the bursts occurring from $f=0$ to $f=1$, the probability
of obtaining a burst in which $n$ fibres break (i.e.\ `size' $n$) is  \cite{hansen94}:
\[
p(n) \simeq n^{-5/2}
\]
When looking at the same distribution for a given interval of load $f\pm \delta f$ one obtains \cite{hansen94}:
\begin{equation}
\label{eq:avalanches}
p(n) \simeq n^{-3/2} \exp(-n/n_{co})
\end{equation}
with
\begin{equation}
\label{eq:co}
n_{co} \approx 1/(1-f)^\gamma ; \, \, \gamma=1.
\end{equation}
That is, a characteristic local scale exists when the system is far from the global rupture, which
however diverges algebraically as the final failure is approached. This behaviour is shown for
the original, mean field,  FBM in figure \ref{fig7} a),
where $p(n)$ and $n_{co}$ (inset) are fitted according to (\ref{eq:avalanches}) and (\ref{eq:co}),
and indicates that the whole fracture process can be viewed as an approach to a critical point,
where $\lim_{f\to 1} n_{co} = \infty$ with a divergence exponent $\gamma = 1$,
and avalanches of any size are possible.  Similar behaviour is
observed in the CFBM \cite{kim05} but not in the LFBM, where the distribution observed is stretched
exponential \cite{zhang96,kloster97}. Critical behaviour in two dimension is only observed for when 
sharing is long range enough \cite{hidalgo02}.

\begin{figure}
\begin{center}
\includegraphics[height=13cm,angle=270]{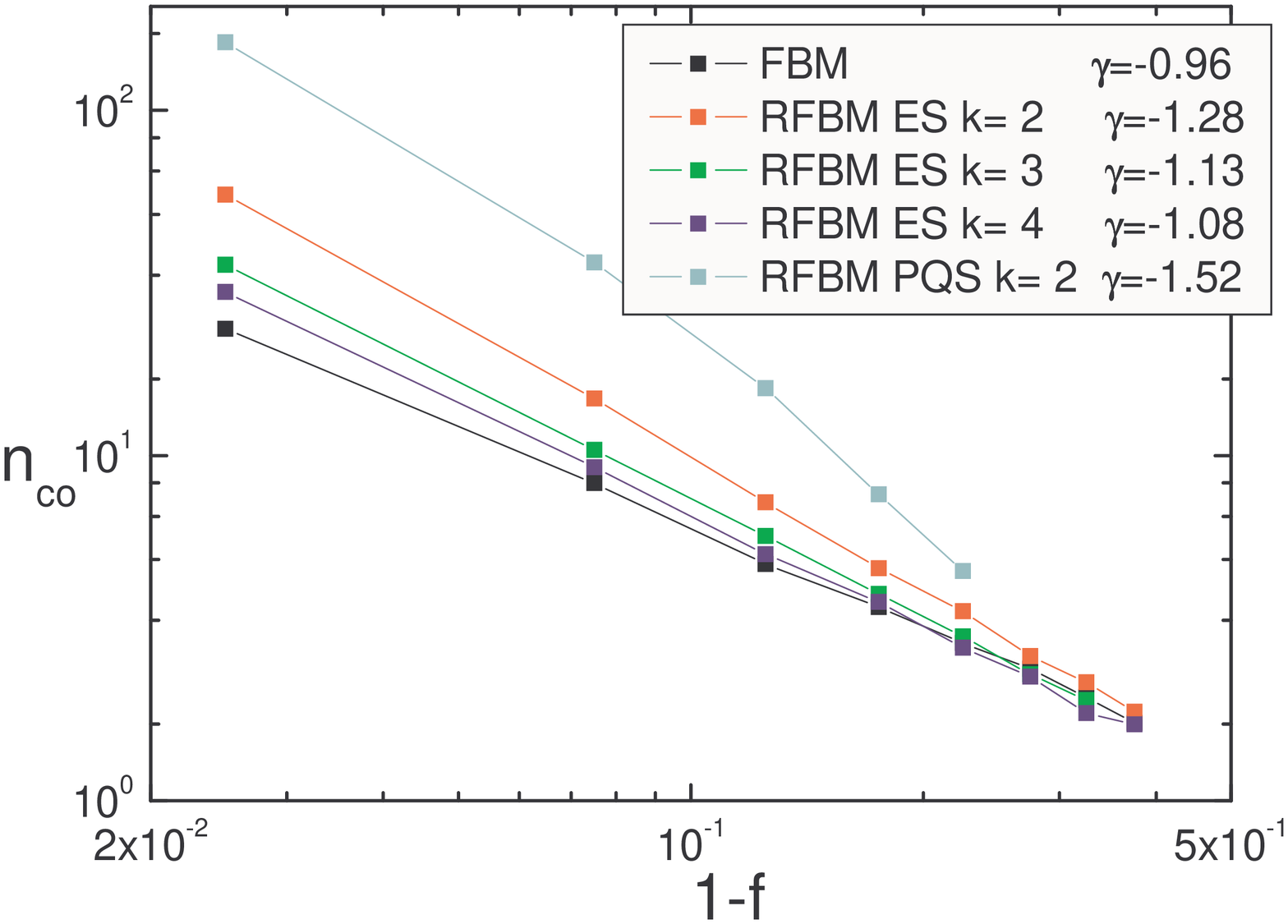}
\end{center}
\caption{\label{fig8}Divergence of the avalanche cut-off $n_{co}$ with $f$ according to (\ref{eq:co}) for
the FBM, the RFBM ES $k=2,3,4$, and for the RFBM PQS $k=2$.
For increasing $k$ the critical exponent $\gamma$ approaches the mean-field FBM value of $\gamma=1$, whereas the
PQS displays $\gamma=1.5$.}
\end{figure}

In figure~\ref{fig7} b) the evolution of the avalanche
distribution for the ES RFBM with $k=2$ is demonstrated.  As in the FBM, the initial avalanches
are of course small  but the distribution broadens as $f$ increases.  Curves are successfully
fitted by  (\ref{eq:avalanches}), indicating behaviour identical to the FBM.
The inset displays the divergence of the cut-off $n_{co}$ as $f \rightarrow 1$ together
with a fit with (\ref{eq:co}), again identical to the FBM model, but
with a different exponent: $\gamma=1.28$. The same critical behaviour is observed for the
RFBM with different values of $k$. For instance, $\gamma=1.08$ for $k=4$. In fact, the exponent ruling the cut-off divergence
is seen to approach
 that of the FBM as $k$ increases, as expected.  This is illustrated by figure \ref{fig8}, where the evolution of the parameter $n_{co}$ is shown as a function of $1-f$ for the FBM,  the ES RFBM with $k=2,3,4$ and  the PQS RFBM (with $k=2$).
In this last case  $\gamma$ has a very high value  $\gamma=1.5$.
For ES the system critical behaviour rapidly approaches that of the FBM \cite{hansen94} and CFBM \cite{kim05} as $k$ increases.
Critical behaviour is not observed in the LFBM with dimension $ = 1 $; on the other hand, CFBM
and the model presented here utilise
a quenched and an annealed random neighbour selection respectively, and so may be considered to have a high dimensionality.
This agrees with the observation that even LFBM in dimensions $\geq 2$ also show criticality~\cite{hidalgo02}.

\section{Summary and discussion}
\label{sec:summary}

We have introduced a model for yielding (RFBM) specifically inspired by the yield of a sheared granular bed under increasing stress.
In the model, when a  force chain fails it discharges its load on $k$ other randomly chosen chains. Each force chain can
withstand a finite amount of stress extracted from some probability distribution. In this respect the model is closely
related to the Fibre Bundle Model (FBM) but employing a novel, random load-sharing rule.
This model may be considered a schematization of any general system in which some external load or
pressure is borne by a number of units, each independent of the others except when
failure triggers some load-transfer process.

For a finite number of units $N$, the yield strength at which global failure occurs displays a probability
distribution with a log-normal shape, a feature common to many systems. This property is
shown to be  largely independent of the choice of units' capacities distribution $t(s)$. For increasing $N$ the
distribution exhibits scaling and finally shrinks to a non-zero deterministic value, at variance with the LFBM (local sharing rule).
This implies a residual strength for the systems in the limit $N  \rightarrow \infty$, a property that we attribute to
the lack of spatial correlations.

Under increasing stress and soon after the loading starts, the distribution of internal loads  evolves to an exponential,
a feature characteristic of  granular media and that also is found to be very robust with respect to the choice
of the capacities distribution $t(s)$. In addition it is also essentially  independent of the random
redistribution rule for broken units.
Considering then the evident  analogies with the $q$-model, we
hypothesize that random neighbour selection is a condition for the emergence
of the exponential load distribution.
This is in stark contrast to the power-law internal load distribution obtained from the 1D LFBM.
Considering also the results relating to random networks, we also conclude that
the absence of spatial correlations gives rise
to the critical fluctuations that the model exhibits as the global breakdown is approached.

\ack
F.D.\ gratefully acknowledges financial support from E.U.\ Nest/Pathfinder project TRIGS under contract
no.\ NEST-2005-PATH-COM-043386.

\appendix
\section*{Appendix}
List of probability densities for the threshold distribustions employed in the examples and related parameters.
\begin{itemize}
 \item[1)] uniform:\\
 $t(s)=1 \, \, \textrm{for} \, \, s \in [0,1)$, $t(s)=0$ otherwise;
\item[2)] exponential:\\
 $ t(s) = e^{-s} \, \,  \textrm{for}  \, \, s \ge 0$;
\item[3)] Gumbel:\\
 $t(s) = e^{-(s+e^{-s})} \, \, \textrm{for} \, \, s \ge 0$;
\item[4)] Weibull:\\
 $t(s) = 2se^{-s^2} \, \, \textrm{for} \, \, s \ge 0$.
\end{itemize}

\noappendix

\section*{References}

\bibliographystyle{unsrt}
\bibliography{FBM-refs,sheargrbibv2}

\newpage

\end{document}